\documentclass[12pt]{article}

\usepackage{amssymb}
\usepackage{amsmath}

\usepackage{amsbsy}

\usepackage{rotating}

\usepackage[usenames]{color} 
\usepackage{colortbl} 

\usepackage{graphics}
\usepackage{epsfig}

\begin{document}%



\title{\bf Bound on a diffuse flux of ultra-high energy neutrinos in the ADD model}

\author{
M.O.~Astashenkov\thanks{Electronic address: mixa.astash@yandex.ru}
\\
{\small Department of Physics, Lomonosov Moscow State University,} \\
{\small 119991 Moscow, Russian Federation}
\\
A.V. Kisselev\thanks{Electronic address: alexandre.kisselev@ihep.ru}
\\
{\small A.A.~Logunov Institute for High Energy Physics, NRC
``Kurchatov Institute'',} \\
{\small 142281 Protvino, Russian Federation} }

\date{}

\maketitle

\thispagestyle{empty}

\bigskip

\begin{abstract}
The search for ultra-high energy downward-going and Earth-skim\-ming
cosmic neutrinos by the Surface Detector array of the Pierre Auger
Observatory (PAO) is analyzed in the ADD model with $n$ extra flat
spatial dimensions. We assumed that the diffuse neutrino flux
$dN_\nu/dE_\nu$ is equal to $k E_\nu^{-2}$ in the energy range
$10^{17}$ eV -- $2.5 \times 10^{19}$ eV. Taking into account that no
neutrino events where found by the PAO, we have estimated an upper
bound on a value of $k$. It is shown that this bound can be stronger
than the upper bound on $k$ recently obtained by the Pierre Auger
Collaboration, depending on $n$ and (n+4)-dimensional gravity scale
$M_D$.
\end{abstract}



\section{Introduction}
\label{sec:intr}

Ultra-high energy (UHE) cosmic neutrinos plays an important role in
particle physics and astrophysics. They help us to determine the
composition of UHE cosmic rays, as well as their origin. In
particular, the detection of UHE neutrino candidates by the Pierre
Auger Observatory (PAO) in coincidence with gravitational wave (GW)
events could constrain the position of the source of GW
\cite{Auger_GW}. Measuring the scattering of UHE cosmic neutrinos
off atmospheric nucleons can probe a new physics that could modify
the neutrino-nucleon cross section at energies above $10^{17}$ eV.
The first observation of high-energy astrophysical neutrinos was
done by the IceCube Collaboration in 2014 \cite{IceCube:2014}. It
was found that the neutrino-nucleon cross section agrees with
predictions in the range 6.3 TeV -- 980 TeV \cite{IceCube:2017}.

To detect neutrino events with energies above $10^{17}$ eV, more
powerful cosmic rays facilities such as the PAO \cite{PAO} and
Telescope Array \cite{TA} are needed. Recently, the Pierre Auger
Collaboration reported on searches for downward-going (DG) UHE
neutrinos \cite{Auger:2015}. The DG incline air showers
\cite{Berezinsky:1969}-\cite{Zas:2005} are initiated by cosmic
neutrinos moving with large zenith angle which interact in the
atmosphere near the Surface Detector (SD) array of the PAO. Note
that the background from hadronic showers is very small at $E_\nu >
10^{17}$ eV and negligible above $10^{19}$ eV
\cite{Anchordoqui:2010}. The data were collected by the SD in the
zenith angle bins $60^\circ - 75^\circ$ and $75^\circ - 90^\circ$
for a period which is equivalent of 6.4 years of a complete PAO SD
working continuously.

The PAO also searched for Earth-skimming (ES) air showers
\cite{Bertou:2002}-\cite{Feng:2002} induced by upward tau neutrinos
at zenith angles $90^\circ - 95^\circ$ which interact in the Earth
producing tau leptons. In their turn, the tau leptons escape the
Earth and initiate showers close to the SD.

No neutrino candidates were found. Assuming the diffuse flux of UHE
neutrinos to be
\begin{equation}\label{flux_en_dependence}
\frac{dN}{dE_\nu} = k \,E_\nu^{-2}
\end{equation}
in the energy range $1.0\times10^{17}$ eV -- $2.5\times10^{19}$ eV,
the upper single-flavor limit to the diffuse flux of UHE neutrinos
was obtained by the Pierre Auger Collaboration
\begin{equation}\label{Auger_bound}
k < 6.4 \times 10^{-9} \mathrm{\ GeV \ cm^{-2} \ s^{-1} \ sr^{-1}}
\;.
\end{equation}

This bound is  approximately four times less than the Waxman-Bachall
bound on cosmic neutrino production in optically thin sources
\cite{WB:2001}. Some cosmogenic neutrino models with a pure proton
composition injected at the sources were rejected by the Auger limit
\eqref{Auger_bound}. The maximum sensitivity  of the SD of the PAO
lies at the neutrino energies around 1 EeV \cite{Auger:2015}. The
IceCube fit of the diffuse single-flavor astrophysical neutrino flux
\cite{IceCube:2015}, if extrapolated to 1 EeV, would give $E_\nu^2
dN/dE_\nu = 0.3 \times 10^{-9} \mathrm{\ GeV \ cm^{-2} \ s^{-1} \
sr^{-1}}$.

The calculations of the exposure of the SD array of the PAO were
done under assumption that neutrino-nucleon collisions in the
atmosphere are described by the SM interactions (in CC and NC
channels).

The goal of the present paper is to estimate the single-flavor bound
on the diffuse flux of UHE cosmic neutrinos in the model with extra
dimensions. Namely, the ADD model \cite{Arkani-Hamed:98} with $n$
extra flat spatial dimensions will be considered. We will assume
that neutrino energy spectrum is of the form $E_\nu^{-2}$
\eqref{flux_en_dependence} in the range $10^{17}$ eV -- $2.5 \times
10^{19}$ eV.



\section{Space-time with large extra dimensions (the ADD model)}
\label{sec:ADD}

Let us briefly remind readers the main features of the ADD model.
The large extra dimensions scenario was postulated in
refs.~\cite{Arkani-Hamed:98}. Its metric looks like
\begin{equation}\label{ADD_metric}
ds^2 = g_{{\mu \nu}}(x) \, dx^{\mu} \, dx^{\nu} + \eta_{_{ab}} \,
dy^a \, dy^b \;,
\end{equation}
where $\mu,\nu = 0,1,2,3$, $a,b=1, \ldots n$, and $\eta_{_{ab}} =
(-1, \ldots, -1)$. All $n$ extra dimensions are compactified with a
size $R_c$.

There is a hierarchy relation between the fundamental mass scale in
$D=4+n$ dimensions, $M_D$, and 4-dimensional Planck mass,
$M_{\mathrm{Pl}}$,
\begin{equation}\label{hierarchy_relation}
M_{\mathrm{Pl}} = V_n \, M_D^{2+n} \;,
\end{equation}
where $V_n$ is a volume of the compactified dimensions. $V_n = (2\pi
R_c)^n$ if the extra dimensions are of a toroidal form. In order
$M_D$ to be of order one or few TeV, the radius of the extra
dimensions should be large. The compactification scale $R_c^{-1}$
ranges from $10^{-3}\hbox{\rm \,eV}$ to $10\hbox{\rm \,MeV}$ as $n$
runs from 2 to 6.

All SM gauge and matter fields are assumed to be confined to a
$3$-dimensional brane embedded into a $(3+n)$-dimensional space,
while the gravity lives in all $D$-dimensional space-time called
bulk.

In linearized gravity we present $D$-dimensional metric $G_{AB}$ in
the form ($A,B = 0,1, \ldots, 3+n$)
\begin{equation}\label{lenearized_appr}
G_{AB}(x,y) = \eta_{_{AB}} + \frac{2}{M_D^{1+n/2}} \, h_{AB}(x,y)
\;.
\end{equation}
Performing the KK mode expansion of the gravitational field
$h_{_{AB}}(x,y)$, we obtain the graviton interaction Lagrangian
density
\begin{equation}\label{Lagrangian}
\mathcal{L}_{\mathrm{int}}(x) = -\frac{1}{\bar M_{\mathrm{Pl}}} \,
T^{\mu \nu}(x) \sum_{n=0}^\infty h_{\mu \nu}^{(n)}(x) \;,
\end{equation}
where $n$ labels the KK excitation level and ${\bar M}_{\mathrm{Pl}}
= M_{\mathrm{Pl}}/\sqrt{8\pi}$ is a reduced Planck mass. $T^{\mu
\nu}(x)$ is the energy-momentum tensor of the matter on the brane.
The masses of the KK graviton modes $h_{\mu \nu}^{(n)}$ are
\begin{equation}\label{KK_masses}
m_n = \frac{\sqrt{n_a n^a}}{R_c} \, , \quad n_a=(n_1,n_2 \ldots n_n)
\;.
\end{equation}
So, a mass splitting is $\Delta m \sim R_c^{-1}$ and we have an
almost continuous spectrum of the gravitons.

One can see from (\ref{Lagrangian}) that the coupling of both
massless and massive graviton is universal and very small ($\sim
1/{\bar M}_{\mathrm{Pl}}$). Nevertheless, all cross sections with
real and virtual production of the massive KK gravitons are defined
by the gravity scale $M_D$, but not by $\bar M_{\mathrm{Pl}}$.

\section{Neutrino-nucleon cross sections}
\label{sec:nu_N}

We intend to consider ultra-high energies of cosmic neutrino, $E_\nu
> 10^{17}$ eV. It corresponds to a large center-of-mass energy of
the neutrino-proton collision, $\sqrt{s} \gtrsim 14$ TeV. Thus, we
are in a transplanckian region $\sqrt{s} \gg M_D$. At the
transplanckian energies a scattering is described by classical
physics~\cite{Giudice:02} -\cite{Emparan:02}, provide an impact
parameter $b$ is lager than the $D$-dimensional Schwarzschild radius
$R_S$ \cite{Myers:86}
\begin{equation}\label{R_S}
R_S(s) = \frac{1}{\sqrt{\pi}} \frac{1}{M_D} \left[ \frac{8 \Gamma
\left( \frac{n+3}{2}\right) }{n+2} \frac{\sqrt{s}}{M_D}
\right]^{\!\frac{1}{n+1}} .
\end{equation}
$R_S$ as a function of the neutrino energy $E_\nu$ is presented in
tabs.~1-3 in Appendix~A ($s = 2 m_N E_\nu$). The transplanckian
regime corresponds to the conditions
\begin{equation}\label{tarns_Pl_regime}
\sqrt{s} \gg M_D \;, \quad \theta \sim (R_S/b)^{\!n+1} \;,
\end{equation}
where $\theta$ is the scattering angle \cite{Giudice:02}.

In the eikonal approximation \cite{Cheng:69}, which is valid  at
small momentum transfer ($-t \ll s$) the leading part of the
scattering amplitude is obtained by summation of all ladder diagrams
with graviton exchange in the $t$-channel \cite{Giudice:02}
-\cite{Emparan:02}. The tree-level exchange of the $D$-dimensional
graviton gives the following Born amplitude
\begin{equation}\label{A_Born}
A_{\mathrm{Born}}(q^2) = \frac{s^2}{M_D^{n+2}} \int \!\!\frac{d^n
q_n}{t-q_n^2} = \pi^{n/2} \Gamma(1-n/2) \left( \frac{-t}{M_D^2}
\right)^{\!n/2-1} \!\!\left( \frac{s}{M_D^2} \right)^{\!2} ,
\end{equation}
where $q_n$ is the momentum transfer in the extra dimensions.
Summing all loop diagrams leads to the eikonal formula
\begin{equation}\label{A_eik}
A_{\mathrm{eik}}(s,t) = -2is \!\int \!\!d^2b \,e^{iq b} \left[
e^{\chi(b)} - 1 \right] ,
\end{equation}
with the eikonal phase
\begin{equation}\label{eik}
\chi(b) = \frac{1}{2s} \int \!\!\frac{d^2q}{(2\pi)^2} \,e^{-iq b}
A_{\mathrm{Born}}(q^2) \;.
\end{equation}
It has been calculated in \cite{Giudice:02}-\cite{Emparan:02} (see
also \cite{Sessolo:08}) to be
\begin{equation}\label{eik_final}
\chi(b) = \left( \frac{b}{b_c} \right)^{\!\!n} ,
\end{equation}
where
\begin{equation}\label{b_c}
b_c = \left[ \frac{(4\pi)^{n/2-1} s \Gamma(n/2)}{2M_D^{n+2}}
\right]^{\!1/n} .
\end{equation}
An energy dependence of $b_c$ for different values of $n$ and $M_D$
is presented in tabs.~1-3 in Appendix~A.

As a result, the final expression of the eikonal amplitude
\eqref{A_eik} is given by
\begin{equation}\label{A_eik_final}
A_{\mathrm{eik}}(s,t) = 4\pi s \,b_c^2 F_n(b_c q) \;,
\end{equation}
\begin{equation}\label{F_n}
F_n(y) = - i \int\limits_0^\infty \!\!dx x J_0(x y) \left[
e^{ix^{-n}} - 1 \right] ,
\end{equation}
where $x = b/b_c$. The eikonal representation of the scattering
amplitude is a good approximation, provided $b > R_S$
\cite{Giudice:02}-\cite{Emparan:02}.

At UHEs the neutrino interacts essentially with the quarks
(antiquarks) and gluons inside the nucleon. Let us define a fraction
of the neutrino energy transferred to the nucleon
\begin{equation}\label{y_def}
y = \frac{E_\nu - E'_\nu}{E_\nu} = \frac{Q^2}{x s} \;,
\end{equation}
where $E_\nu (E'_\nu)$ is the initial (final) energy of the
neutrino, and $x$ is the fraction of nucleon momentum carried by
parton $i$ ($i = q, \bar{q}, g$). Taking into account above
mentioned formulas, we get the differential neutrino-nucleon cross
section
\begin{equation}\label{dif_cs}
\frac{d^2\sigma}{dxdy} = \pi s \sum_{i}x f_{i}(x,\mu^2) \,
b_{c}^4(\hat s) \,|F_{n}(b_c Q)|^2 \;,
\end{equation}
where $\hat{s} = x s$, and $Q = \sqrt{y \hat{s}}$. The quantities
$f_{i}(x,\mu^2)$ are the parton distribution functions (PDFs).
Following ref.~\cite{Sessolo:08}, we put $\mu^2 = Q^2$. We use the
CT14 set for the PDFs \cite{CT14}. For a chosen value of $n$ we take
$M_D$ to be equal to a 95$\%$ CL lower limit on $M_D$ obtained
recently by the CMS Collaboration (see fig.~11 in
\cite{CM:limits_MD}). For instance, $M_D^{\min} = 2.3$ TeV (2.5 TeV)
for $n=2$ (6). In order to calculate total cross sections, we
integrate \eqref{dif_cs} in the region $Q_0^2 < Q^2 < R_S^{-2}$
\cite{Sessolo:08}. As in \cite{Sessolo:08}, we put $Q_0^2=0.01
m_W^2$, where $m_W$ is the W-boson mass.

As it was mentioned above, the eikonal approximation can be used if
$Q^2 < R_S^{-2}$ ($b > R_S)$. In the rest of integration region $s
\geqslant Q^2 > R_S^{-2}$, that corresponds to the region $b < R_S$
in the impact parameter space, one expects that the neutrino and a
parton inside the nucleon will form a black hole. In such a case,
the cross section can be estimated as
\cite{Feng:02}-\cite{Giddings:02}
\begin{equation}\label{CS_bh}
\sigma_{\nu N \rightarrow \mathrm{BH}}(s) = \pi \sum_{i}
\!\!\int\limits_{(M_{\mathrm{bh}}^{\mathrm{min}})^2/s}^1 \!\!\!\!dx
f_i (x, \bar{\mu}^2) R_S^2(\hat{s}) \;.
\end{equation}
We put $\bar{\mu}^2  = xs$. The dependence of $\sigma_{\nu N
\rightarrow \mathrm{BH}}$ on the choice of $\bar{\mu}^2$ and
$M_{\mathrm{bh}}^{\mathrm{min}}$ is discussed in
\cite{Anchordoque:02}-\cite{Ahn:02}. For chosen $n$, $M_D$, we take
$M_{\mathrm{bh}}^{\mathrm{min}}$ to be equal to the 95$\%$ CL lower
limit on $M_{\mathrm{bh}}$ for the same $n$ and $M_D$ obtained by
the CMS Collaboration \cite{CM:limits_MBH}. As one can see from
fig.~6 in \cite{CM:limits_MBH} and tabs.~1-3 in Appendix~A,
$M_{\mathrm{bh}}^{\mathrm{min}} \gg R_S^{-1}$ for all $E_\nu$, if $2
\leqslant n \leqslant 6$, and $2 \mathrm{\ TeV} < M_D < 6$ TeV.

The black hole production by cosmic rays was studied in a number of
papers (see, for an example, \cite{Feng:02},
\cite{Anchordoque:02}-\cite{Ringwald:02}).
\begin{figure}[htb]
\begin{center}
\includegraphics[width=6cm,clip]{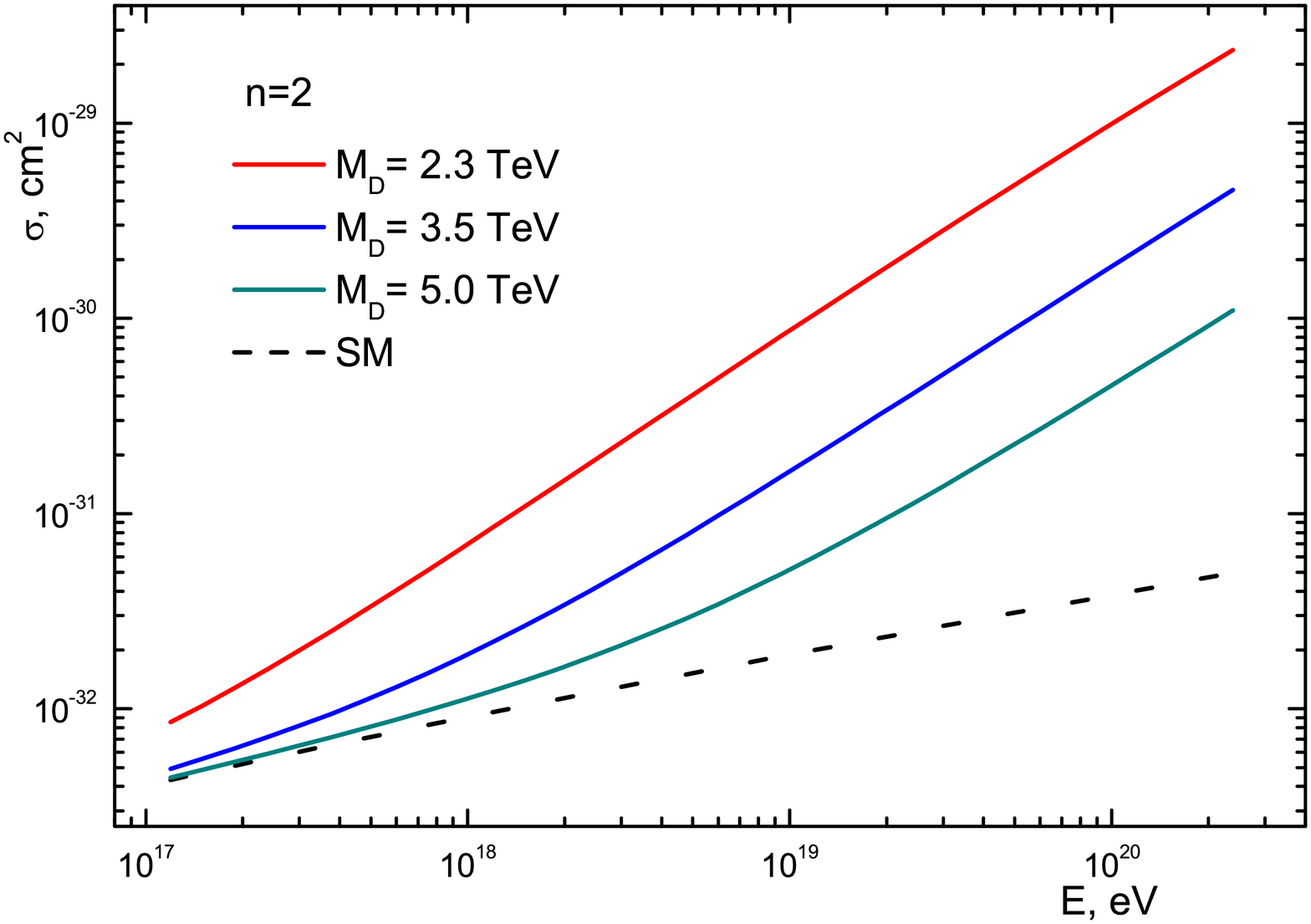} \hspace{.5cm}
\includegraphics[width=6cm,clip]{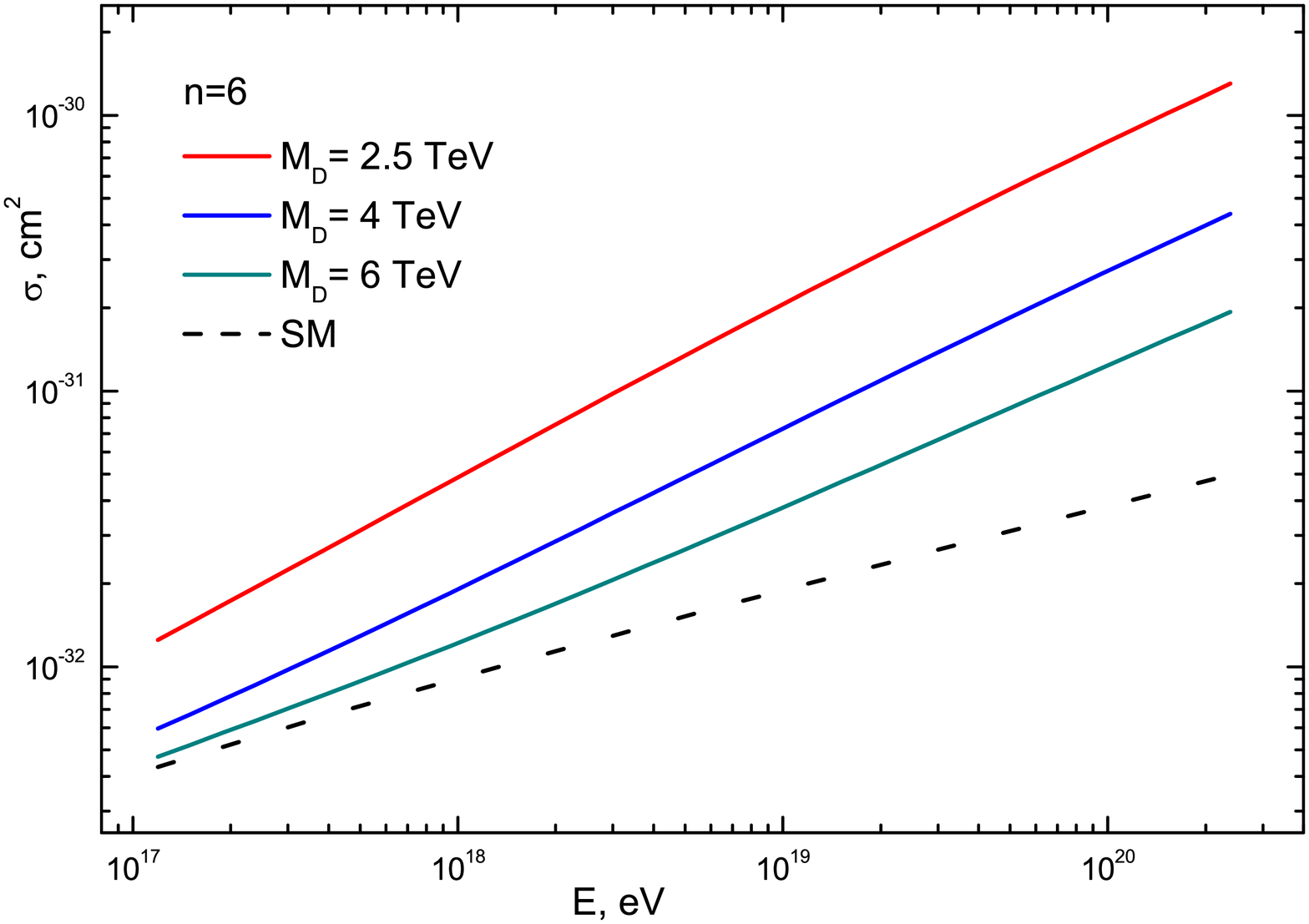}
\caption{Left panel: the neutrino total cross sections for $n=2$ and
$M_D = 2.3$ TeV, 3.5 TeV, 5.0 TeV (solid lines). Right panel: the
same as on the left panel, but for $n=6$ and $M_D = 2.5$ TeV, 4.0
TeV, 6.0 TeV. For comparison, the neutrino CC total cross section is
shown by the dashed lines.} \label{fig:n2sum}
\end{center}
\end{figure}
%
\begin{figure}[htb]
\begin{center}
\includegraphics[width=6cm,clip]{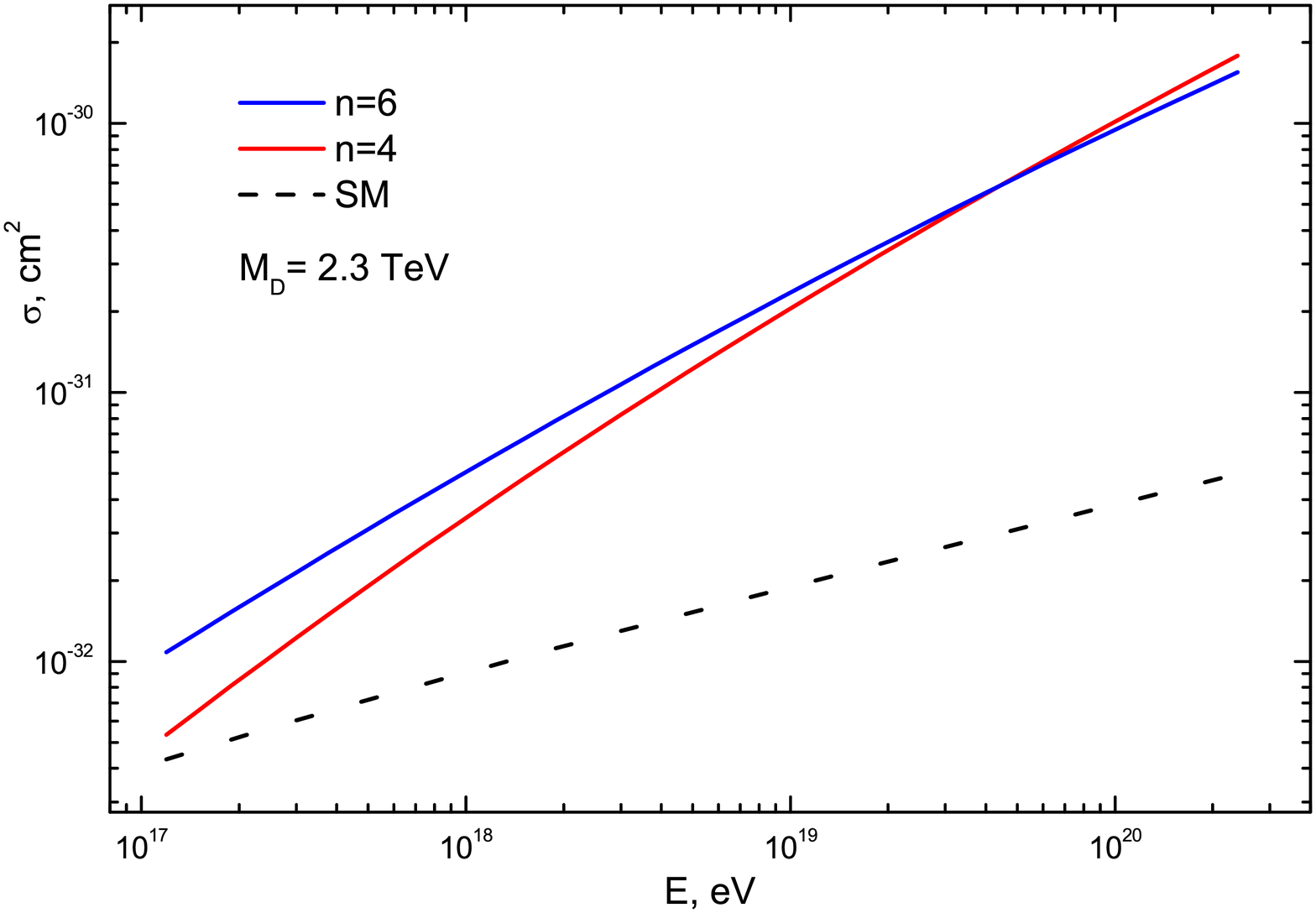} \hspace{.5cm}
\includegraphics[width=6cm,clip]{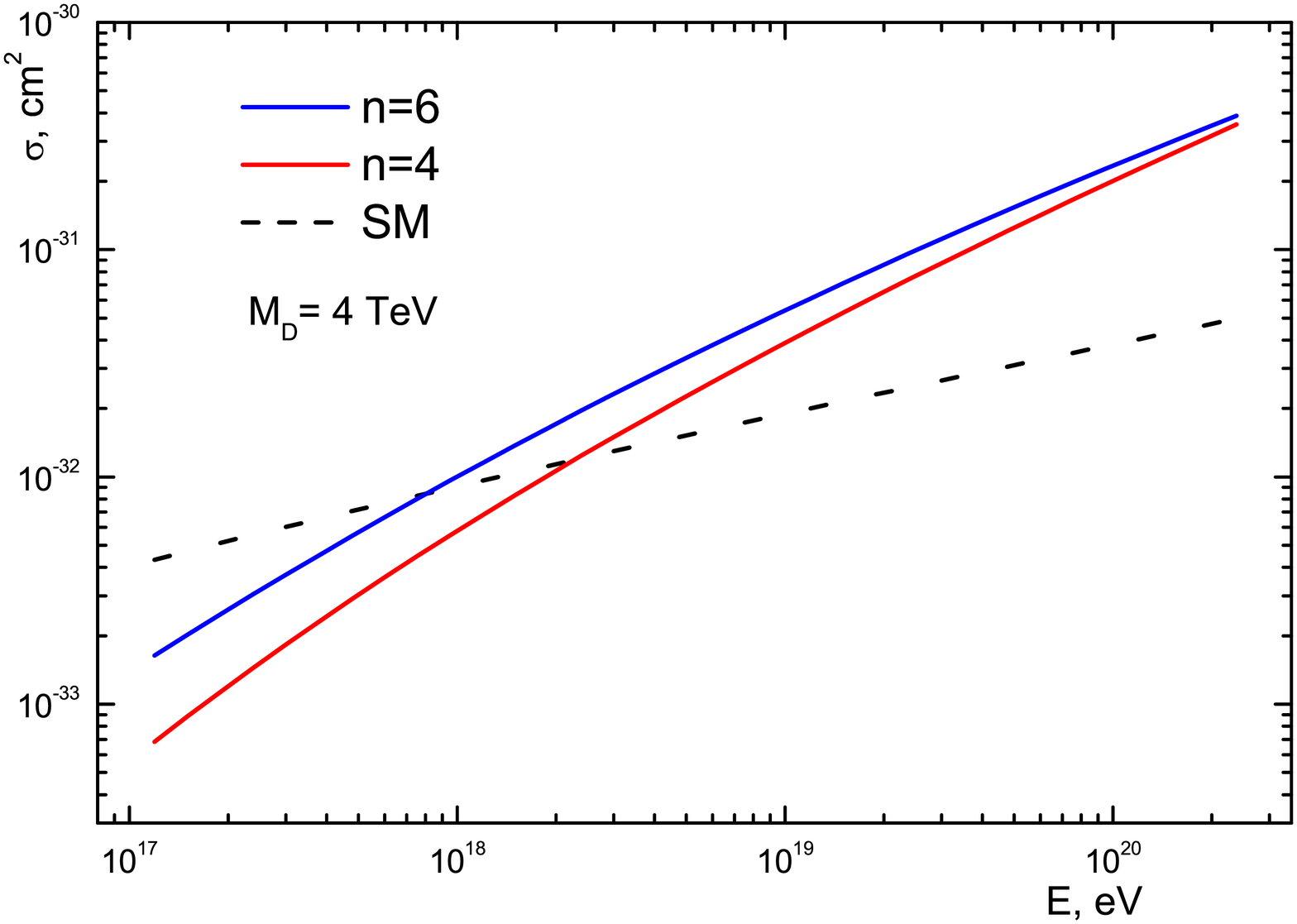}
\caption{Left panel: the neutrino cross sections in the ADD model
for $M_D = 2.3$ TeV and $n = 4$, 6 (solid lines, no SM contribution
is included). Dashed line: the neutrino CC total cross section.
Right panel: the same as on the left panel, but for $M_D = 4$.}
\label{fig:n4n6MD2.3}
\end{center}
\end{figure}

As for the SM neutrino interaction, we adopt the neutrino-nucleon
cross sections in \cite{Sarkar:2008}, since the Pierre Auger
Collaboration \cite{Auger:2015} has obtained  limit
\eqref{Auger_bound} with the use of these SM cross sections.

The total cross sections as functions of the $D$-dimensional mass
scale $M_D$ and number of the extra dimensions $n$ are shown in
figs.~\ref{fig:n2sum}-\ref{fig:n4n6MD2.3}. Let us note that at
$E_\nu
> 10^{19}$ eV the cross section $\sigma_{\nu N \rightarrow
\mathrm{BH}}$ rises with $n$, while the eikonal cross section
decreases. The combined effects of these two factors is that the
difference of the cross sections for $n=4$ and $n=6$ tends to zero
as $E_\nu$ grows (see fig.~\ref{fig:n4n6MD2.3}).

Our calculations of the cross sections is not an end in itself but
it will enable us to estimate exposures for both DG and ES neutrino
events at the SD array of the PAO in the ADD model and thus to put
limits on the diffuse single-flavor flux of UHE neutrinos.

\section{Limits on diffuse flux of UHE neutrinos in the ADD model}
\label{sec:limits}

In \cite{Kisselev:2016} the following functional dependence of the
DG event rate on the new physics cross section
$\sigma_{\mathrm{NP}}$ was proposed for UHE neutrino events
\begin{equation}\label{DG:BSM_vs_SM}
\mathcal{E}_{\mathrm{BSM}}^{\mathrm{DG}} (E_\nu) =
\mathcal{E}_{\mathrm{SM}}^{\mathrm{DG}} (E_\nu)\,
\frac{\sigma_{\mathrm{SM}}^{\mathrm{eff}}(E_\nu) +
\sigma_{\mathrm{NP}}(E_\nu)}{\sigma_{\mathrm{SM}}^{\mathrm{eff}}(E_\nu)}
\;,
\end{equation}
where $\mathcal{E}_{\mathrm{BSM}}^{\mathrm{DG}}$
($\mathcal{E}_{\mathrm{SM}}^{\mathrm{DG}}$) is the exposure of the
SD of the PAO with (without) account of the new interaction. In
addition, instead of $\sigma_{\mathrm{CC}}$, an effective SM cross
section $\sigma_{\mathrm{SM}}^{\mathrm{eff}}$ is introduce in
\eqref{DG:BSM_vs_SM}:
\begin{equation}\label{sigma_eff}
\sigma_{\mathrm{SM}}^{\mathrm{eff}} = \sigma_{\mathrm{CC}}
\!\!\sum_{i=e,\mu,\tau} \! \, m_{\mathrm{CC}}^i +
3\sigma_{\mathrm{NC}} \, m_{\mathrm{NC}} + \sigma_{\mathrm{CC}}\,
m_{\mathrm{mount}} \;.
\end{equation}
Here $m_{\mathrm{CC}}^i$ and $m_{\mathrm{NC}}$ are relative mass
apertures for charged current (CC) and neutral current (NC)
interactions of the DG neutrinos at the PAO. The mass aperture
$m_{\mathrm{mount}}$ corresponds to the CC interaction of a $\tau$
neutrino within the mountains around the PAO. The relative mass
apertures as functions of the neutrino energy where calculated using
the data in Table~I of ref.~\cite{Auger:2011}. Note that
$\sum_{i=e,\mu,\tau} m_{\mathrm{CC}}^i + 3 m_{\mathrm{NC}} +
m_{\mathrm{mount}} = 1$.

\begin{figure}[htb]
\centering
\includegraphics[width=8cm,clip]{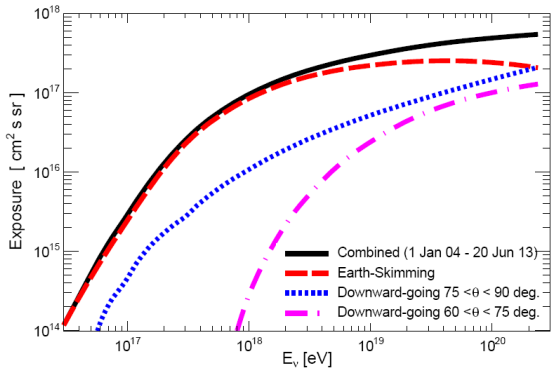}
\caption{The combined exposure of the SD array of the PAO (1 January
2004-20 June 2013) as a function of the neutrino energy. The
individual exposures are also shown (fig.~3 from
ref.~\cite{Auger:2015}).} \label{fig:Auger_exposures}
\end{figure}

In contrast to the DG neutrino exposure, the exposure of the ES
neutrinos decreases with the rise of the neutrino total cross
section \cite{Kisselev:2016}
\begin{equation}\label{ES:BSM_vs_SM}
\mathcal{E}_{\mathrm{BSM}}^{\mathrm{ES}}(E_\nu) =
\mathcal{E}_{\mathrm{SM}}^{\mathrm{ES}}(E_\nu) \,
\frac{\sigma_{\mathrm{CC}}^2(E_\nu)}{[\sigma_{\mathrm{CC}}(E_\nu) +
\sigma_{\mathrm{NP}}(E_\nu)]^2} \;.
\end{equation}

\begin{figure}[htb]
\centering
\includegraphics[width=6cm,clip]{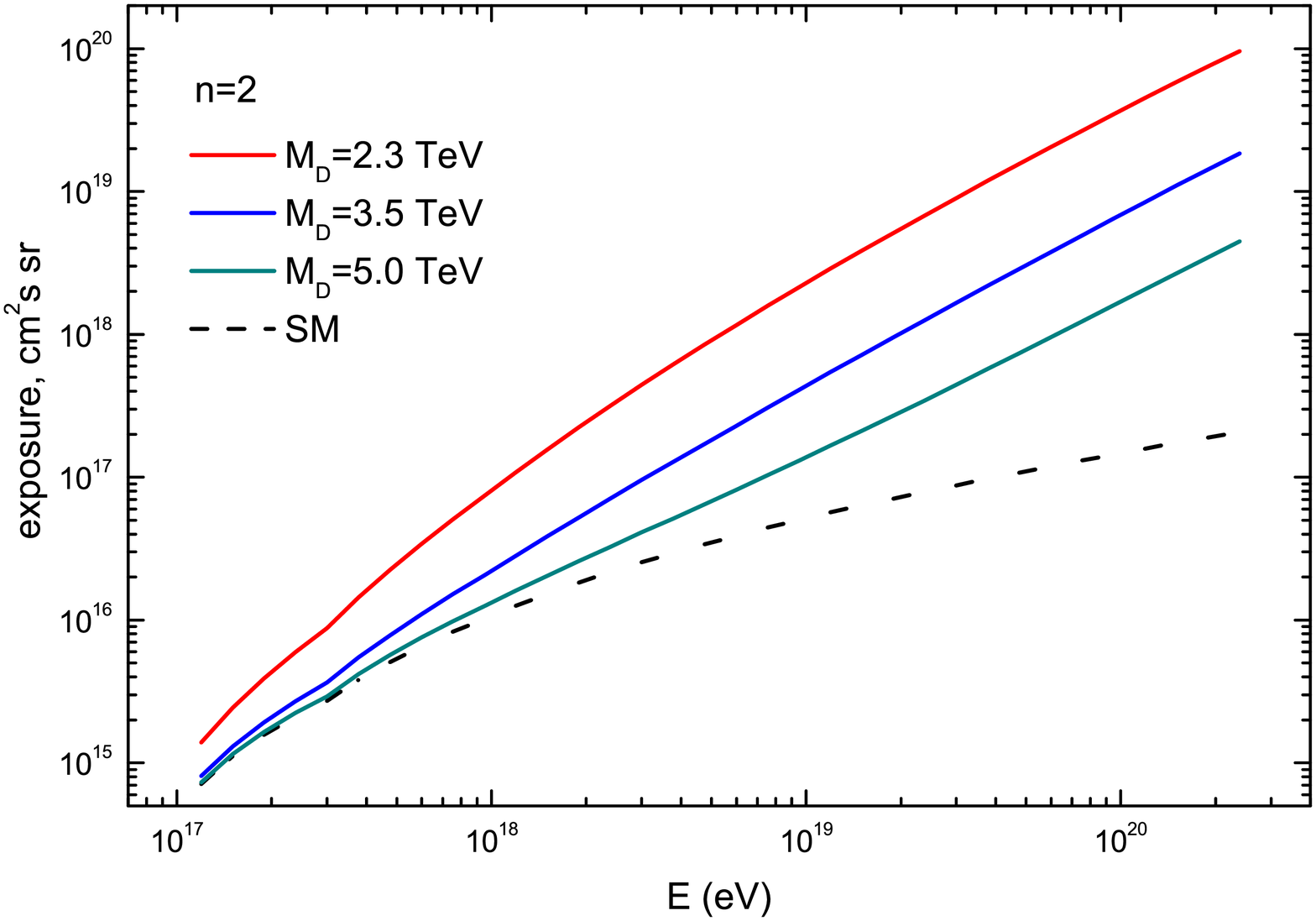} \hspace{5mm}
\includegraphics[width=6cm,clip]{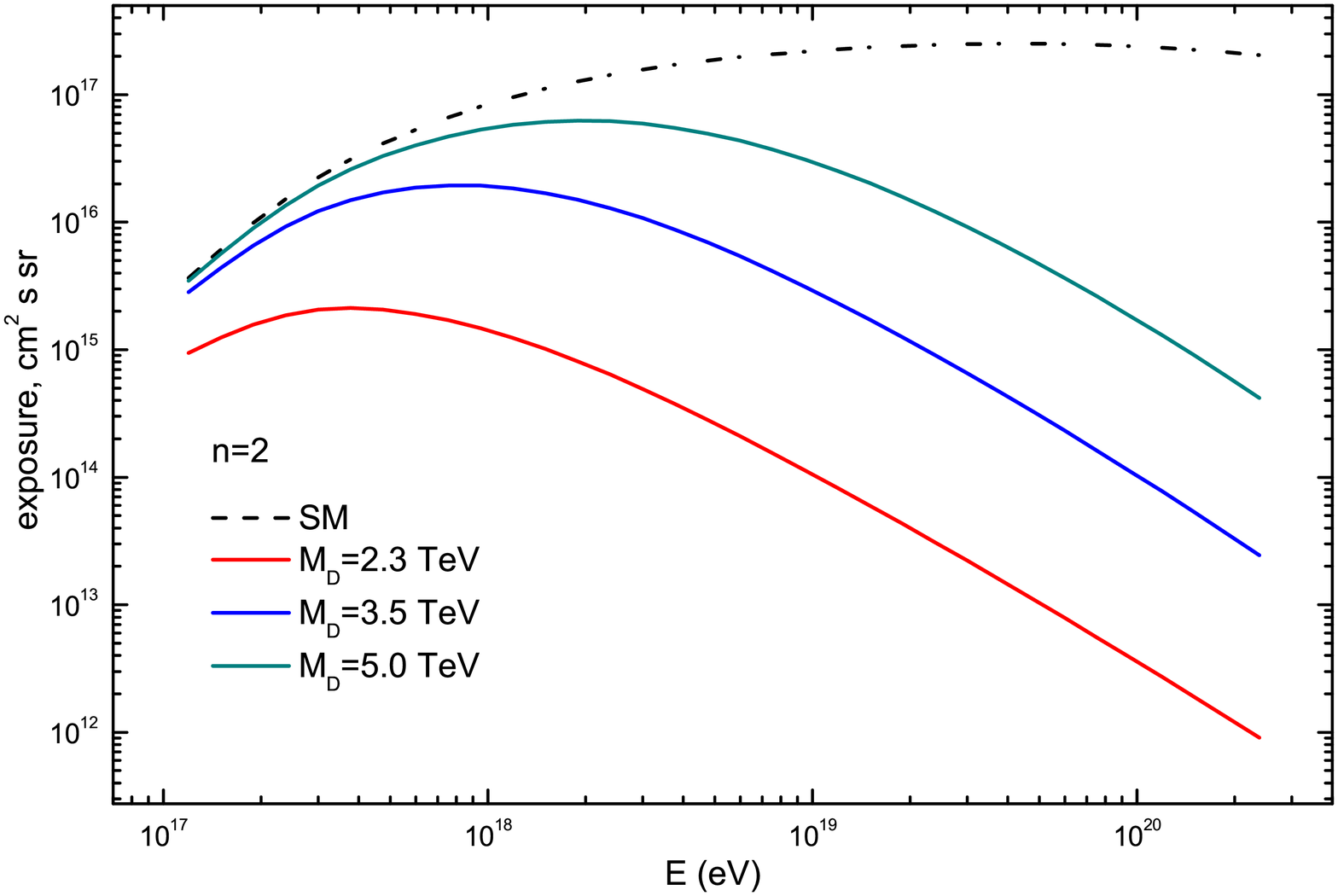}
\caption{Left panel: the expected exposures of the SD array of the
PAO for the DG neutrinos with zenith angle $75^\circ < \theta <
90^\circ$ in the ADD model. Right panel: the expected exposures of
the SD array of the PAO for the ES neutrinos in the ADD model.}
\label{fig:n2_exp}
\end{figure}

\begin{figure}[htb]
\centering
\includegraphics[width=6cm,clip]{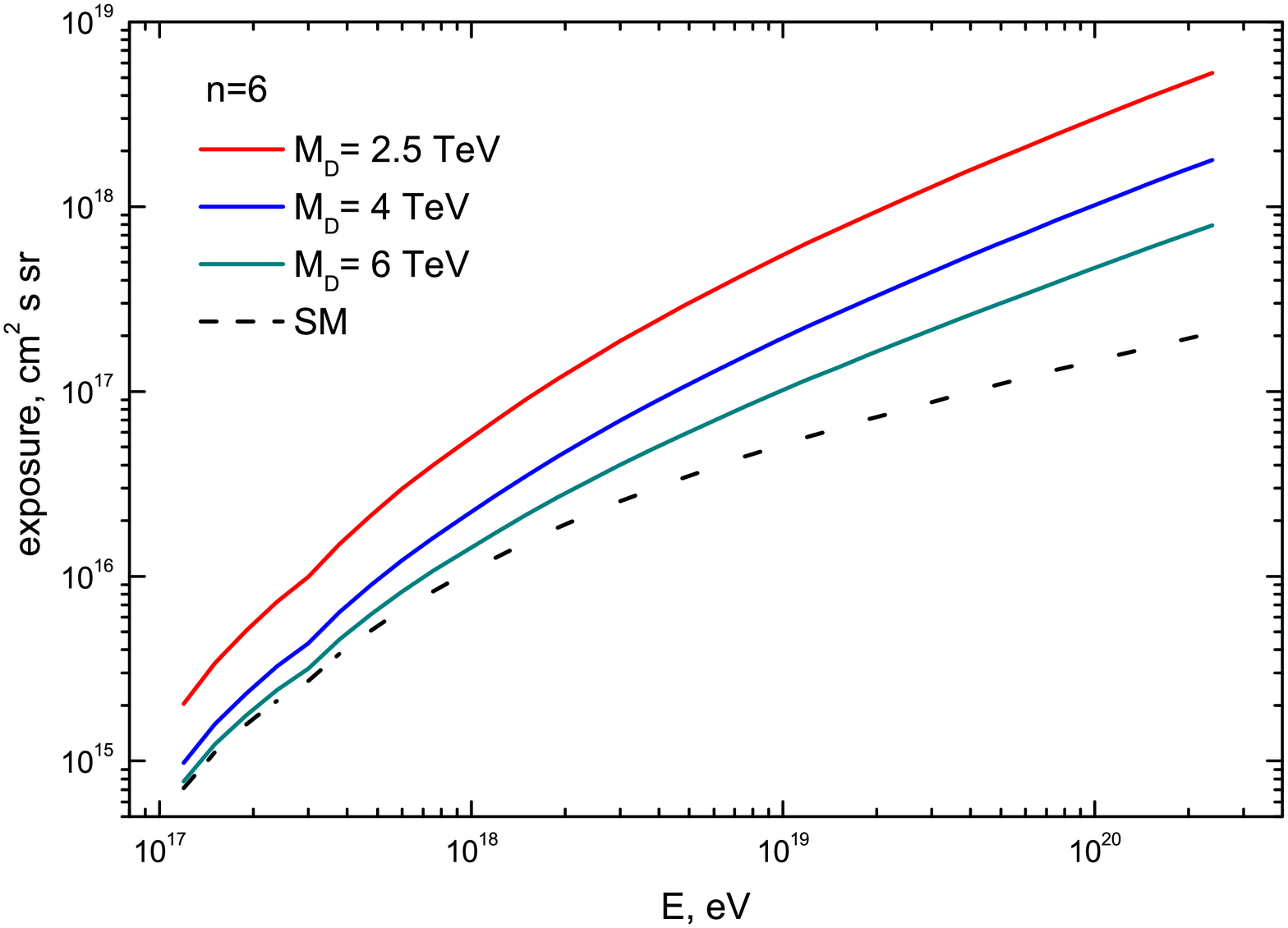} \hspace{5mm}
\includegraphics[width=6cm,clip]{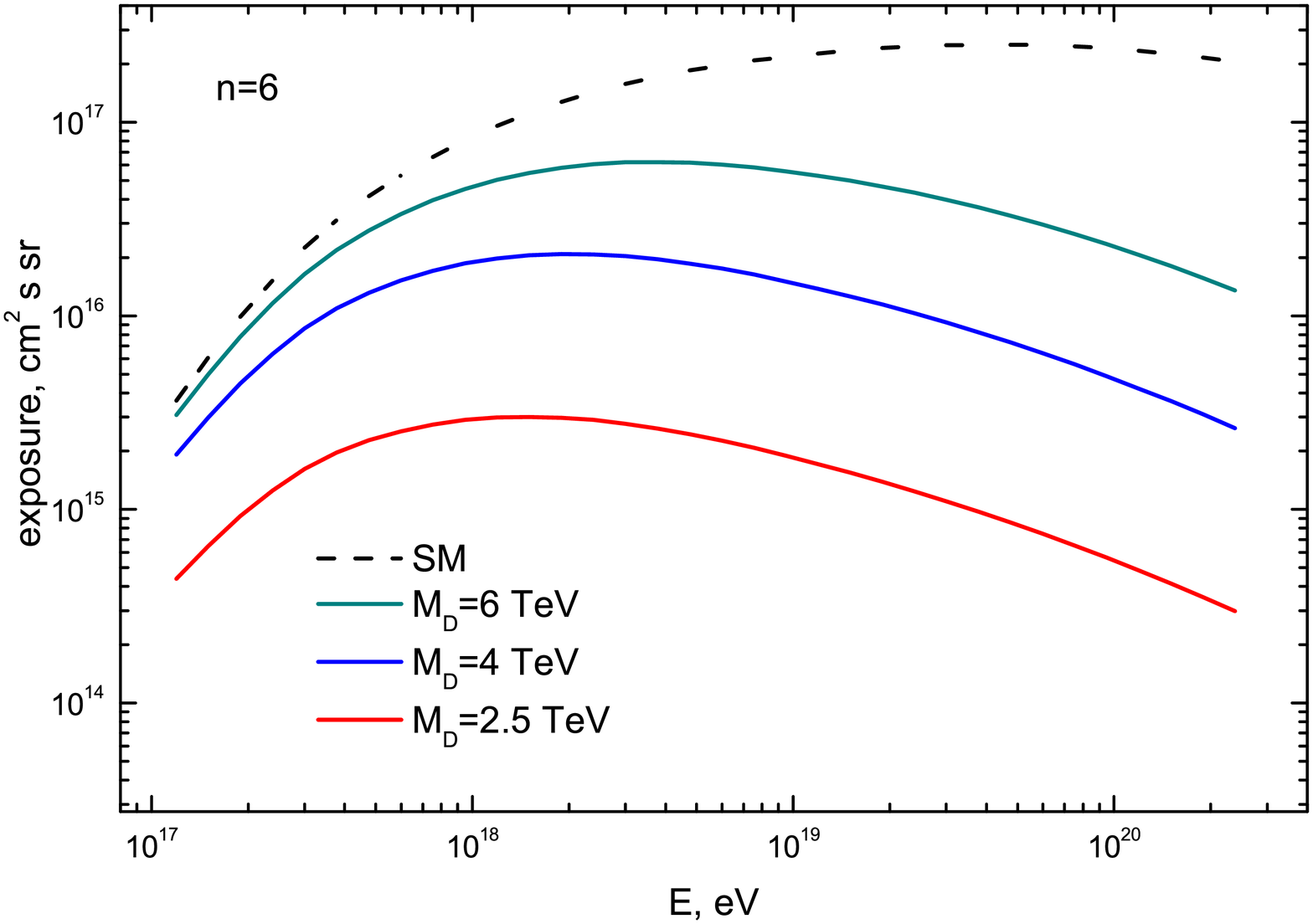}
\caption{The same as in fig.~\ref{fig:n2_exp}, but for $n=6$.}
\label{fig:n6_exp}
\end{figure}

The formulas \eqref{DG:BSM_vs_SM} and \eqref{ES:BSM_vs_SM} allowed
us to calculate exposures of the SD of the PAO for the period 1
January 2004 -- 20 June 2013 expected in the ADD model. The PAO data
on the exposures for the SM neutrino interactions in the region from
$\log(E_\nu/\mathrm{eV}) = 17$ to 20.5 were used (see
fig.~\ref{fig:Auger_exposures} taken from ref.~\cite{Auger:2015}).
The results of our calculations are presented in
figs.~\ref{fig:n2_exp}-\ref{fig:n6_exp}.

%
\begin{figure}[htb]
\centering
\includegraphics[width=6cm,clip]{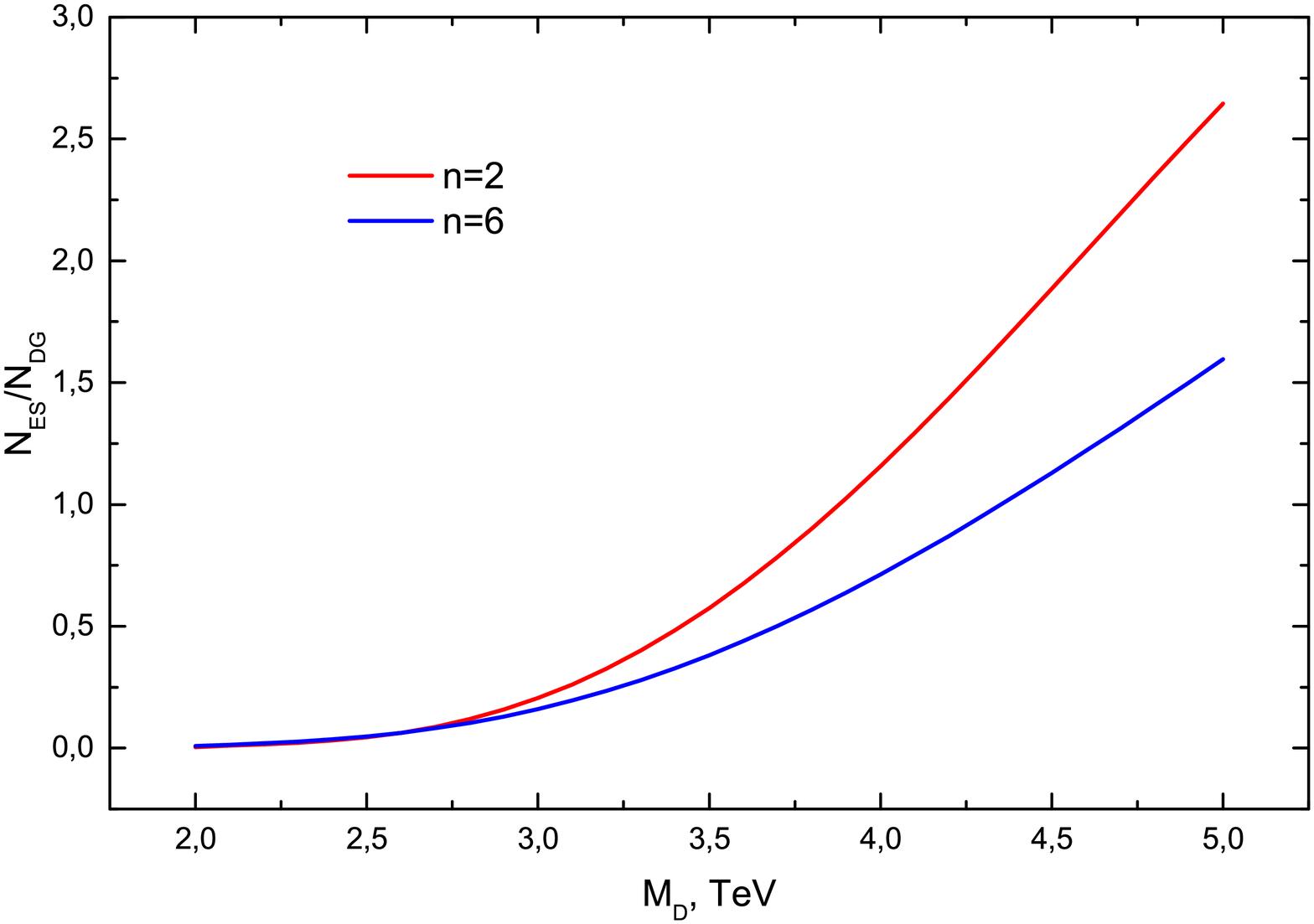}
\caption{The expected ratio of the ES neutrinos to the DG neutrinos
with zenith angle $75^\circ < \theta < 90^\circ$ at the SD array of
the PAO as a function of the gravity scale $M_D$ for two values of
$n$.} \label{fig:ES_DG}
\end{figure}

We assume that the astrophysical flux arrives isotropically from all
directions, and neutrino flavor composition is $\nu_e : \nu_\mu
:\nu_\tau  = 1 : 1 : 1$. Following Pierre Auger Collaboration, we
also assume that the flux is described by a power law of the form
\eqref{flux_en_dependence}. Then the upper limit on the value of $k$
can be estimated as \cite{Auger:2015}
\begin{equation}\label{k_int}
k = \frac{N_{\mathrm{up}}}{\int \!E_\nu^{-2}
\mathcal{E}_{\mathrm{tot}}(E_\nu) d E_\nu} \;,
\end{equation}
where $N_{\mathrm{up}}$ is an actual value of the upper limit on the
signal events which depends on the number of the observed events and
total exposure
\begin{equation}\label{exp_tot}
\mathcal{E}_{\mathrm{tot}} =
\mathcal{E}_{\mathrm{BSM}}^{\mathrm{DG}} +
\mathcal{E}_{\mathrm{BSM}}^{\mathrm{ES}} \;,
\end{equation}
see eqs.~\eqref{DG:BSM_vs_SM} and \eqref{ES:BSM_vs_SM}. Since the
PAO sees no events, we put $N_{\mathrm{up}} = 2.39$, assuming a
number of expected background events to be zero \cite{Auger:2015}.

As one can see in fig.~\ref{fig:n2sum}, in the ADD model the cross
sections rise more rapidly with the neutrino energy than the SM
cross sections. As a result, the exposure for the DG events,
$\mathcal{E}_{\mathrm{BSM}}^{\mathrm{DG}}$ \eqref{DG:BSM_vs_SM},
rises, while the exposure for the ES events,
$\mathcal{E}_{\mathrm{BSM}}^{\mathrm{ES}}$ \eqref{ES:BSM_vs_SM},
decreases as $E_\nu$ grows (see figs.~\ref{fig:n2_exp},
\ref{fig:n6_exp}). The expected ratio of the ES neutrinos to the DG
neutrinos with zenith angle $75^\circ < \theta < 90^\circ$ is shown
in fig.~\ref{fig:ES_DG}.

\begin{figure}[htb]
\centering
\includegraphics[width=6cm,clip]{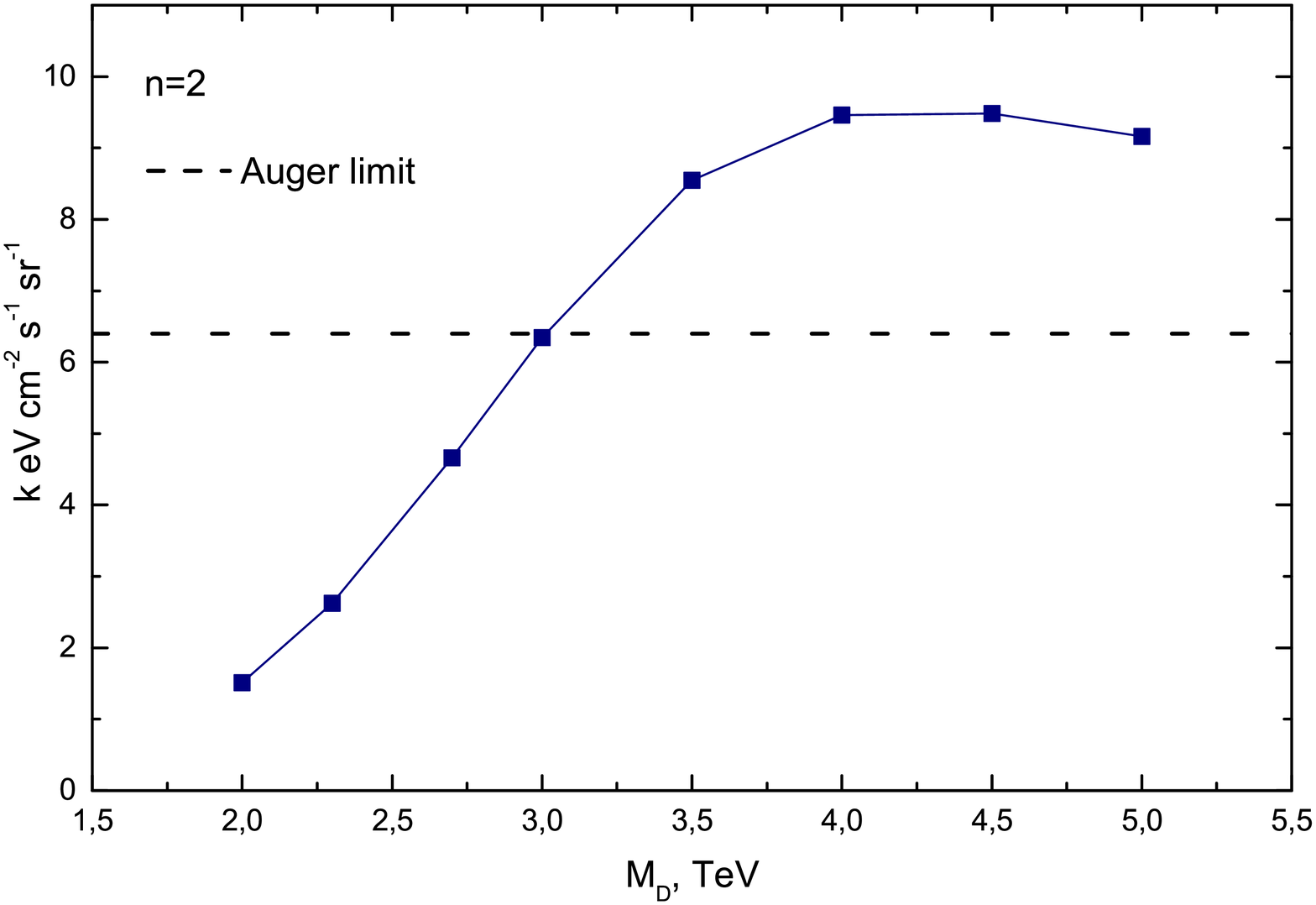} \hspace{5mm}
\includegraphics[width=6cm,clip]{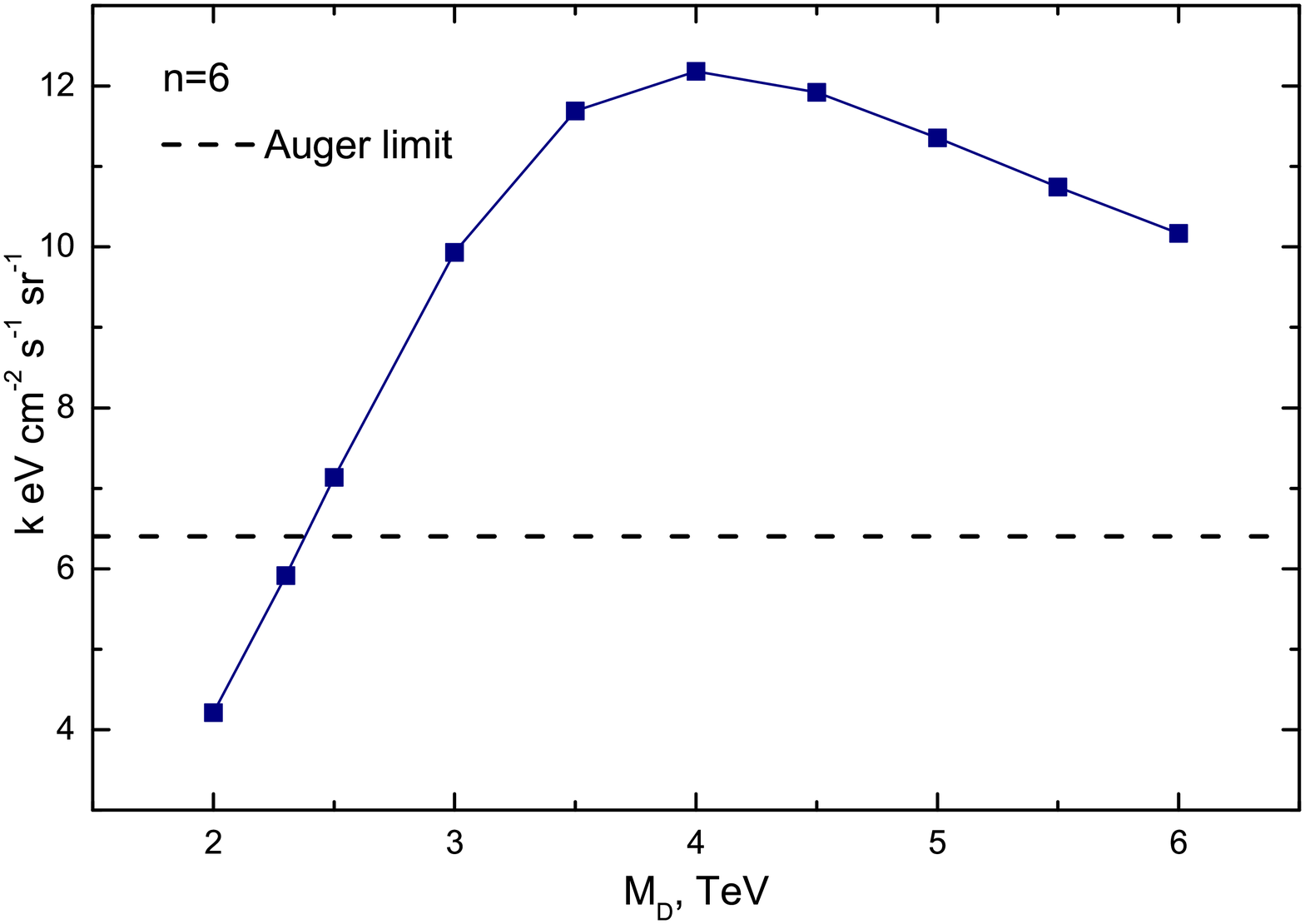}
\caption{Left panel: the upper bound on the value of $k$ as a
function of D-dimensional Planck scale $M_D$ for $n=2$ in the ADD
model. Dashed line is the PAO upper limit \cite{Auger:2015}. Right
panel: the same as on the left panel, but for $n=6$.}
\label{fig:k_n_MD}
\end{figure}

\begin{figure}[htb]
\centering
\includegraphics[width=6cm,clip]{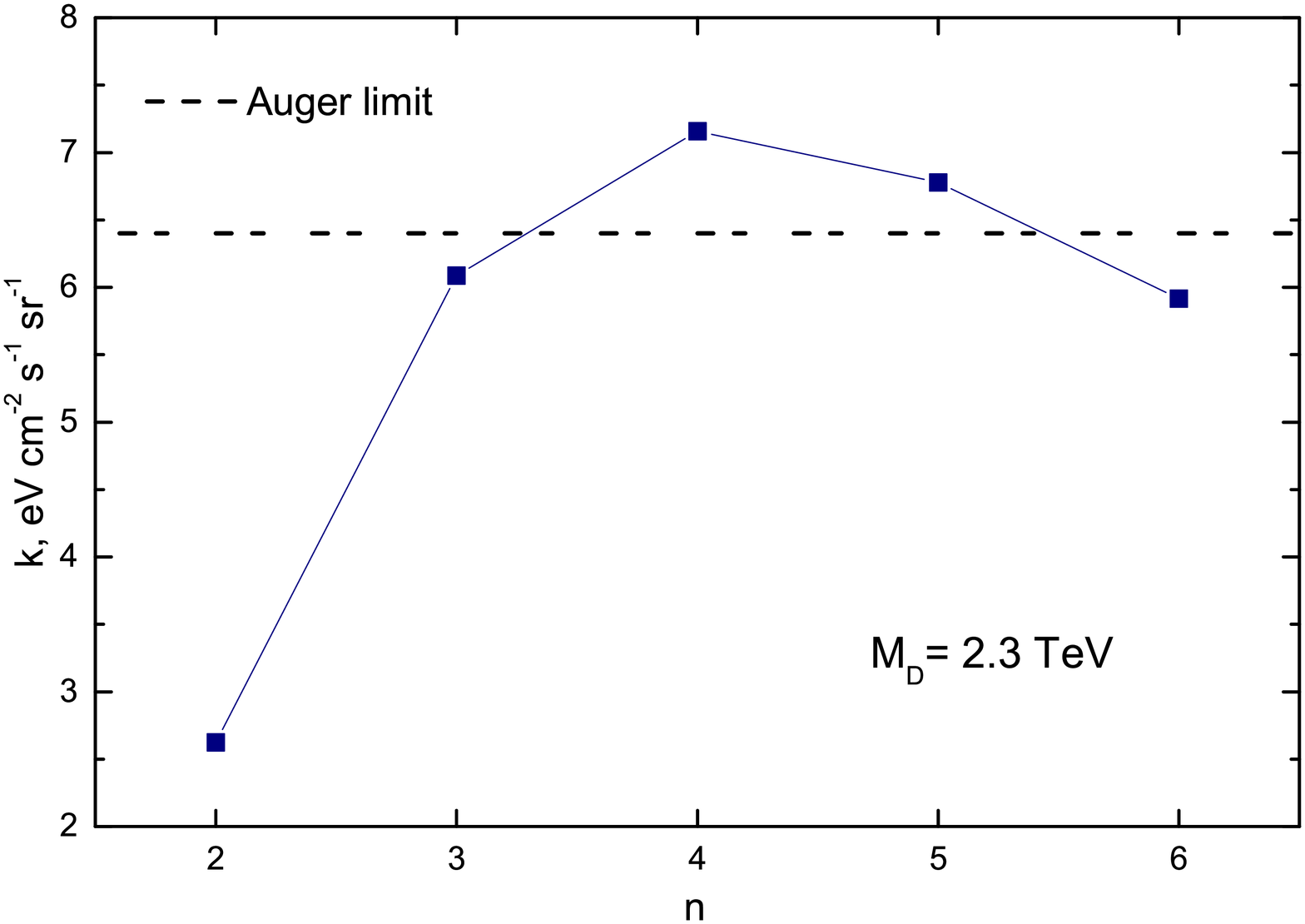} \hspace{5mm}
\includegraphics[width=6cm,clip]{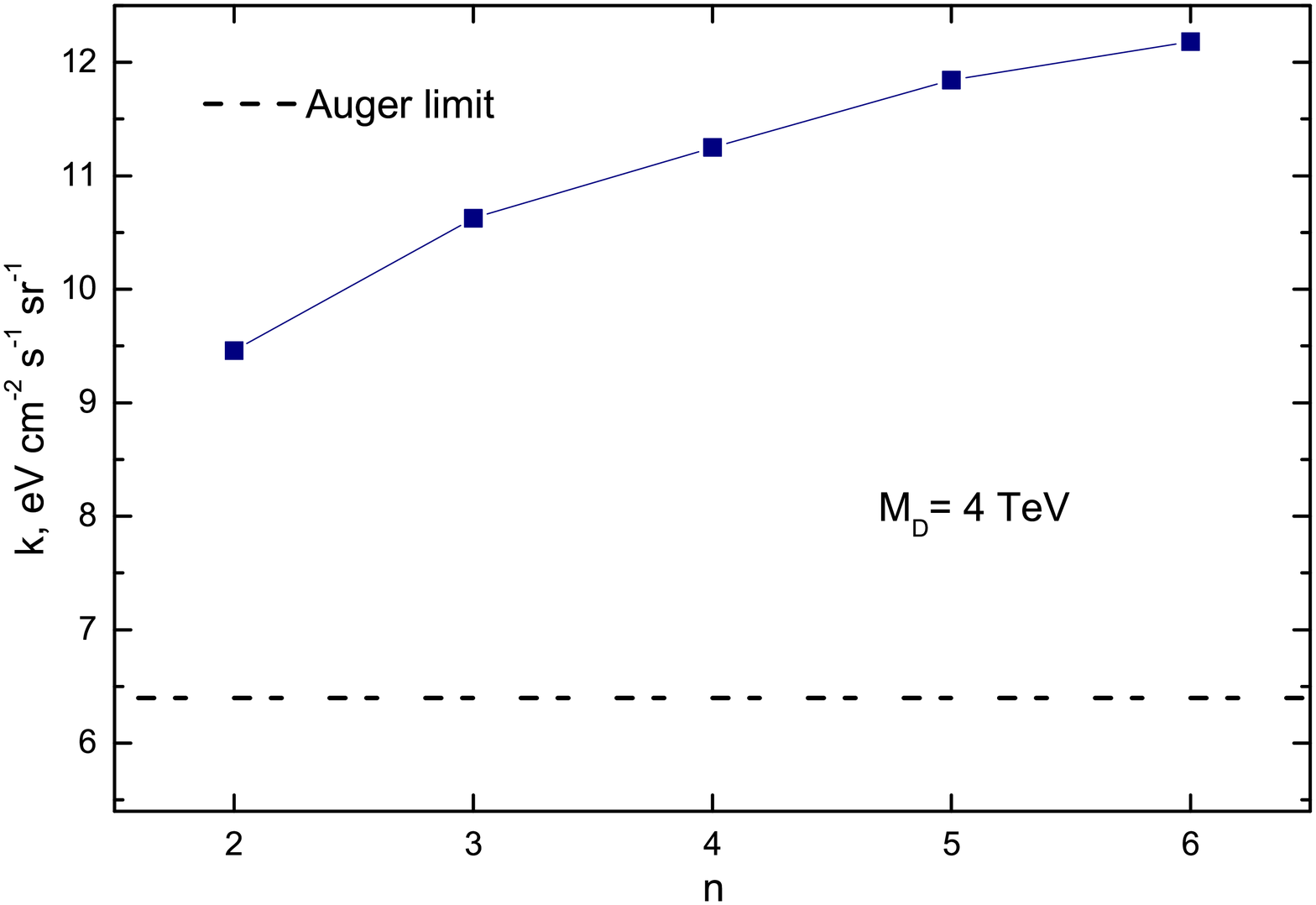}
\caption{Left panel: the upper bound on the value of $k$ as a
function of number of extra dimensions $n$ for $M_D= 2.3$ TeV. Right
panel: the same as on the left panel, but for $M_D= 4.0$ TeV.}
\label{fig:k_MD_n}
\end{figure}

As a result, for some values of $n$ and $M_D$, the total expected
exposure in the ADD model \eqref{exp_tot} can be larger than the
Auger exposure calculated on the assumption that the
neutrino-nucleon scattering is defined by the SM interactions only.
Correspondingly, an upper bound on $k$ defined by eq.~\eqref{k_int}
can be even stronger than the bound obtained by the Pierre Auger
Collaboration \eqref{Auger_bound}. It is demonstrated by
figs.~\ref{fig:k_n_MD}-\ref{fig:k_MD_n}, in which the PAO upper
bound on the value of $k$ is also shown.

%


\section{Conclusions}

Using the exposure of the PAO for the period equivalent of 6.4 years
of the complete PAO SD array working continuously, we have estimated
the exposures for the neutrino induced events expected in the
scenario with the large flat extra dimensions of the space-time.
Both downward-going and Earth-skimming UHE cosmic neutrinos are
considered.

The exposures are defined by the neutrino-nucleon cross sections in
the ADD model. In the transplanckian region and large impact
parameters $b > R_S$ the eikonal approximation is valid. In such a
case, the scattering amplitude is given by the exchanges of the
$t$-channel massive gravitons. At small $b < R_S$, the eikonal
approximation breaks down, and the production of the black holes is
assumed. The dependence of the exposures on the number of extra
dimensions $n$ and the gravity scale $M_D$ is obtained
(figs.~\ref{fig:n2_exp}-\ref{fig:n6_exp}).

Our main goal was to calculate the single-flavor upper limit on the
diffuse neutrino flux in the presence of the massive graviton
interactions in the ADD model. We assumed that the flux of UHE
neutrinos is proportional to $E_\nu^{-2}$
\eqref{flux_en_dependence}. Our results demonstrate us that in the
ADD model the upper bound on the diffuse neutrino flux can be
stronger that the PAO limit \eqref{Auger_bound}, depending on the
parameter of the ADD model, $n$ and $M_D$. As one can see in
fig.~\ref{fig:k_n_MD}, it takes place for $M_D < 3.01$ TeV (2.38
TeV), if $n=2$ (6). For $M_D = 2.3$ TeV it is true for $n \leqslant
3$ and $n \geqslant 6$ (left panel of fig.~\ref{fig:k_MD_n}).
However, with the increase of $M_D$ our bound becomes weaker than
the PAO bound for all $n$ (right panel of fig.~\ref{fig:k_MD_n}).

It can be understood as follows. Remember that the upper limit on
the neutrino diffuse flux \eqref{flux_en_dependence} is given by
formula \eqref{k_int}. In the presence of the extra dimensions, the
neutrino-nucleon cross section grows with the neutrino energy more
rapidly than the SM one (fig.~\ref{fig:n2_exp}). Correspondingly,
the expected exposure for the DG neutrino events,
$\mathcal{E}_{\mathrm{BSM}}^{\mathrm{DG}}$ \eqref{DG:BSM_vs_SM},
also rises. On the contrary, the exposure for the ES neutrino
events, $\mathcal{E}_{\mathrm{BSM}}^{\mathrm{ES}}$
\eqref{ES:BSM_vs_SM}, decreases as the energy grows
(figs.~\ref{fig:n2_exp}-\ref{fig:n6_exp}). As a result, the total
expected exposure of the SD array of the PAO,
$\mathcal{E}_{\mathrm{tot}}$ \eqref{exp_tot}, may be larger than the
total exposure obtained by the Pierre Auger Collaboration
(fig.~\ref{fig:Auger_exposures}), provided that the integrated
increase of
$E_\nu^{-2}\mathcal{E}_{\mathrm{BSM}}^{\mathrm{DG}}(E_\nu)$ prevails
over the integrated reduction of
$E_\nu^{-2}\mathcal{E}_{\mathrm{BSM}}^{\mathrm{ES}}(E_\nu)$. As
$M_D$ grows, the upper limit on the value of $k$ tends to the PAO
limit \eqref{Auger_bound} from above, as one can see on the right
panel of fig.~\ref{fig:k_n_MD}.



\section*{Acknowledgements}

The authors are indebted to J. Alvarez-Mu\~{n}iz for sending us the
numerical data on the exposures of the SD array of the PAO.



\setcounter{equation}{0}
\renewcommand{\theequation}{A.\arabic{equation}}

\section*{Appendix A}
\label{app:A}

Here we present an energy dependence of the parameter $b_c$
\eqref{b_c} and $D$-dimensional Schwarzschild radius squared $R_S^2$
\eqref{R_S} for different values of the number of the extra
dimensions $n$ and $D$-dimensional gravity scale $M_D$.
\bigskip

\noindent Table~1. The parameter $b_c$ and Schwarzschild radius
squared $R_S^2$ for $n=2$, $M_D = 2.3$ TeV as a function of the
neutrino energy $E_\nu$.
\bigskip
\begin{center}
\begin{tabular}{|c|c|c|}
  \hline
$E_\nu$, eV & $b_c$, GeV$^{-1}$ & $R_S^2$,  GeV$^{-2}$ \\
  \hline \hline
$1.00000000 \cdot 10^{17}$ & $1.831113\cdot 10^{-3}$ & $ 5.155385\cdot 10^{-7}$ \\
$1.50356136 \cdot 10^{17}$ & $2.245307\cdot 10^{-3}$ & $ 5.906109\cdot 10^{-7}$ \\
$1.00000000 \cdot 10^{18}$ & $5.790488\cdot 10^{-3}$ & $ 1.110694\cdot 10^{-6}$ \\
$1.50356136 \cdot 10^{18}$ & $7.100285\cdot 10^{-3}$ & $ 1.272433\cdot 10^{-6}$ \\
$1.00000000 \cdot 10^{19}$ & $1.831113\cdot 10^{-2}$ & $ 2.392918\cdot 10^{-6}$ \\
$1.50356136 \cdot 10^{19}$ & $2.245307\cdot 10^{-2}$ & $ 2.741373\cdot 10^{-6}$ \\
$1.00000000 \cdot 10^{20}$ & $5.790488\cdot 10^{-2}$ & $ 5.155385\cdot 10^{-6}$ \\
$1.50356136 \cdot 10^{20}$ & $7.100285\cdot 10^{-2}$ & $ 5.906109\cdot 10^{-6}$ \\
$2.38298316 \cdot 10^{20}$ & $8.938727\cdot 10^{-2}$ & $ 6.886017\cdot 10^{-6}$ \\
  \hline
\end{tabular}
\end{center}
\bigskip

\noindent Table~2. The same as in tab.~1, but for $n=4$.
\bigskip
\begin{center}
\begin{tabular}{|c|c|c|}
  \hline
$E_\nu$, eV & $b_c$, GeV$^{-1}$ & $R_S^2$,  GeV$^{-2}$ \\
  \hline \hline
$1.00000000 \cdot 10^{17}$ & $1.679949\cdot 10^{-3}$ & $ 1.161700\cdot 10^{-6}$ \\
$1.50356136 \cdot 10^{17}$ & $1.860272\cdot 10^{-3}$ & $ 1.260429\cdot 10^{-6}$  \\
$1.00000000 \cdot 10^{18}$ & $2.987419\cdot 10^{-3}$ & $ 1.841171\cdot 10^{-6}$ \\
$1.50356136 \cdot 10^{18}$ & $3.308083\cdot 10^{-3}$ & $ 1.997645\cdot 10^{-6}$ \\
$1.00000000 \cdot 10^{19}$ & $5.312466\cdot 10^{-3}$ & $ 2.918059\cdot 10^{-6}$ \\
$1.50356136 \cdot 10^{19}$ & $5.882696\cdot 10^{-3}$ & $ 3.166054\cdot 10^{-6}$ \\
$1.00000000 \cdot 10^{20}$ & $9.447048\cdot 10^{-3}$ & $ 4.624811\cdot 10^{-6}$ \\
$1.50356136 \cdot 10^{20}$ & $1.046108\cdot 10^{-2}$ & $ 5.017857\cdot 10^{-6}$ \\
$2.38298316 \cdot 10^{20}$ & $1.173752\cdot 10^{-2}$ & $ 5.501970\cdot 10^{-6}$ \\
  \hline
\end{tabular}
\end{center}
\bigskip

\noindent Table~3. The same as in tab.~1, but for $n=6$, $M_D = 2.5$
TeV.
\bigskip
\begin{center}
\begin{tabular}{|c|c|c|}
  \hline
$E_\nu$, eV & $b_c$, GeV$^{-1}$ & $R_S^2$,  GeV$^{-2}$ \\
  \hline \hline
$1.00000000 \cdot 10^{17}$ & $1.639502\cdot 10^{-3}$ & $ 1.550374\cdot 10^{-6}$ \\
$1.50356136 \cdot 10^{17}$ & $1.754818\cdot 10^{-3}$ & $ 1.643386\cdot 10^{-6}$ \\
$1.00000000 \cdot 10^{18}$ & $2.406460\cdot 10^{-3}$ & $ 2.154238\cdot 10^{-6}$ \\
$1.50356136 \cdot 10^{18}$ & $2.575721\cdot 10^{-3}$ & $ 2.283477\cdot 10^{-6}$ \\
$1.00000000 \cdot 10^{19}$ & $3.532200\cdot 10^{-3}$ & $ 2.993304\cdot 10^{-6}$ \\
$1.50356136 \cdot 10^{19}$ & $3.780641\cdot 10^{-3}$ & $ 3.172881\cdot 10^{-6}$ \\
$1.00000000 \cdot 10^{20}$ & $5.184560\cdot 10^{-3}$ & $ 4.159182\cdot 10^{-6}$ \\
$1.50356136 \cdot 10^{20}$ & $5.549222\cdot 10^{-3}$ & $ 4.408704\cdot 10^{-6}$ \\
$2.38298316 \cdot 10^{20}$ & $5.991912\cdot 10^{-3}$ & $ 4.708497\cdot 10^{-6}$ \\
  \hline
\end{tabular}
\end{center}
\bigskip

Note that the invariant energy squared of the neutrino-nucleon
scattering is equal to $s = 2 m_N E_\nu$.




\end{document}